\documentclass[prd,twocolumn,nofootinbib,showpacs,amsmath,amssymb]{revtex4}

\usepackage{graphicx}
\usepackage{dcolumn}
\usepackage{bm}

\begin{document}

\title{Elastic $pp$ and $\bar pp$ scattering\\ in the models of
unitarized pomeron}

\author{E. Martynov}

\affiliation{%
Bogolyubov Institute
for Theoretical Physics, 03680 Kiev, Ukraine.\\
e-mail: martynov@bitp.kiev.ua
}%

\date{\today}

\begin{abstract}
Elastic scattering amplitudes dominated by the Pomeron singularity which obey the principal unitarity bounds at high energies are constructed and analyzed. Confronting the models of double and triple (at $t=0$) Pomeron pole (supplemented by some terms responsible for the low energy behaviour) with existing experimental data on  $pp$ and $\bar pp$ total and differential cross sections at $\sqrt{s}\geq 5$ GeV and $|t|\leq 6$ GeV$^{2}$ we are able to tune the form of the Pomeron singularity. Actually the good agreement with those data is received for both models though the behaviour given by the dipole model is more preferable in some aspects. The predictions made for the LHC energy values display, however, the quite noticeable difference between the predictions of models at $t\approx -0.4$ GeV$^{2}$. Apparently the future results of TOTEM will be more conclusive to make a true choice.
\end{abstract}

\maketitle

\section{\label{sec:level1}Introduction}

The forthcoming TOTEM experiment at LHC will provide us, in fact,
with the first measurements of soft pomeron (strictly speaking pomeron
and odderon) as the contributions of secondary reggeons are negligible at such high energies. Then obviously the precise measurement of $pp$ differential cross section makes it possible to discriminate the various pomeron models comparing their particular predictions. Certainly, such an analysis makes sense only if the same data set is used with the model parameters reliably fixed. For the time being there are three model types for elastic hadron scattering amplitudes which reproduce rising cross sections experimentally measured with a high precision.

\begin{itemize}
\item  Models treating Pomeron (and odderon as well) as a simple pole in a
complex momentum plane located righter of unity, $\alpha_{P}(0)=1+\varepsilon\approx 1.1$ \cite{PVL,DGP}. In order to describe a dip-bump structure in differential cross section one should take into account the cuts in one or another form. Such a pomeron violates unitarity bound $\sigma_{t}(s)\leq Cln^{2}s$ at $s\to \infty$. However, the argument that unitarity corrections are important only at higher energies justifies this approach.
\item Pomeron with $\alpha_{P}(0)>1$ is an input in some scheme of
unitarization (for example, eikonal or quasieikonal \cite{eik}, $U$-matrix
models \cite{UM}). Having done the unitarization all such models give
$\sigma_{t}(s)\propto ln^2{s}$, whereas other characteristic predictions
depend on the concrete model.
\item Another way to construct amplitude is just to take into account
unitarity and analytical requirements from the beginning  as well as
experimental information on the cross sections (e.g. growth of total
cross sections). Such a model we named in what follows as model of unitarized pomeron. Here most successful examples are tripole pomeron
($\sigma_{t}(s)\propto ln^{2}s$) \cite{LNGLN,GLN,AGN} and dipole pomeron
($\sigma_{t}(s)\propto lns$) \cite{JMS,DGMP}.
\end{itemize}

Within the third approach we consider the models of tripole and dipole pomeron. These models are most successful in a description of all data on the forward scattering data \cite{COMPETE}.

As it was shown \cite{COMPETE,DGMP2} the total cross sections of meson and nucleon interactions are described with the minimal $\chi^{2}$ in the dipole and tripole models in which forward scattering amplitudes are parameterized in explicit analytic form. This conclusion was confirmed by analysis applying the dispersion relations for real part of amplitudes \cite{clms-disp}.

The elastic scattering at small-$|t|$ ($0.1\leq |t|\leq 0.5$ GeV$^{2}$)
($pp, \bar pp, \pi^{\pm}p$ and $K^{\pm}p$) was analyzed in detail \cite{CLM}. The particular model was considered as a combination of hard (with $\alpha_{h}(0)\approx 1.4$) and soft $\alpha_{s}(0)\approx 1.1$ pomeron contributions.  It was noticed in \cite{CLMS} that additional hard pomeron essentially improves the description of the meson and nucleon data on parameter  $\rho=\Re eA(s,0)/\Im mA(s,0)$ comparing with ordinary soft pomeron model.  Extension of the model to higher $|t|$ can be done within some scheme of unitarization (e.g. eikonal, quasieikonal, $U-matrix$)  taking into account pomeron rescatterings or cuts.

Here we focus on the dipole and tripole pomeron models. Without entering the details we note here these models describe the small-$|t|$ differential cross sections with the same level of precision ($\chi^{2}/dof\lesssim 1.05$, dof$\equiv$ degrees of freedom) as the model of \cite{CLM} did. The purpose of the present paper is to demonstrate the description of the data on elastic $pp$ and $\bar pp$ scattering at low and middle $t$ in the dipole and tripole pomeron models.

In Sec. \ref{sec:Gen.Con} we remind the general restrictions on hardness of pomeron singularity and form of its trajectory at small $t$, imposed by unitarity bounds on cross sections. In Sec. \ref{sec:Dpar} and in Sec. \ref{sec:Tpar} parametrizations of $pp$ and $\bar pp$ elastic scattering
amplitudes are presented dealing with dipole and tripole pomeron models,
correspondingly. Results of least square analysis for both models as well as their comparison are given in the Sec. \ref{sec:expdata}.

\section{General constraints}\label{sec:Gen.Con}
Let us reiterate here that the model with $\sigma_{t}(s)\propto
ln^{2}s$ is not compatible with a linear pomeron trajectory having
the intercept 1. Indeed, let us assume that
$$
\alpha_{P}(t)=1+\alpha_{P}'t
$$
and the partial wave amplitude develops the form
\begin{equation}\label{eq:j-gen}
\varphi (j,t)=\eta(j)\frac{\beta(j,t)}{\left
[j-1-\alpha_{P}'t\right]^{n}}\approx \frac{i\beta(1,t)}{\left
[j-1-\alpha_{P}'t\right]^{n}},\end{equation}
\[ \eta(j)=\frac{1+e^{-i\pi
j}}{-\sin\pi j}.\]

In ($s,t$)-representation amplitude $\varphi(j,t)$ is transformed to
\begin{equation}\label{eq:s,t-gen}
a(s,t)=\frac{1}{2\pi i}\int dj \varphi (j,t)e^{\xi (j-1)}, \quad
\xi=ln(s/s_{0}).
\end{equation}
Then, we have pomeron contribution at large $s$ as
\begin{equation}\label{eq:s,t-gen1}
a(s,t)\approx -g(t)[ln(-is/s_{0})]^{n-1}(-is/s_{0})^{\alpha'_{P}t}
\end{equation}
where
\[
g(t)=\beta (t)/\\sin(\pi\alpha_{P}(t)/2).
\]
If as usually  $g(t)=g\exp(bt)$ then we obtain
\begin{eqnarray}\label{eq:sigma-0}
\sigma_{t}(s)&&\propto ln^{n-1}s,\nonumber\\
\sigma_{el}(s)&&\propto
\frac{1}{s^{2}}\int \limits_{-\infty}^{0}dt|a(s,t)|^{2}\propto
ln^{2n-3}s.
\end{eqnarray}
According to the obvious inequality,
\begin{equation}\label{eq:unitbound}
\sigma_{el}(s)\leq \sigma_{t}(s)
\end{equation}
we have
\begin{equation}\label{eq:nbound}
2n-3\leq n-1 \qquad \Rightarrow \qquad n\leq 2.
\end{equation}

Thus we come to the conclusion that a model with $\sigma_{t}(s)\propto
ln^{2}s$ is incompatible with a linear pomeron trajectory. In other words  the partial amplitude Eq. (\ref{eq:j-gen}) with $n=3$ (but used in some papers) in principle is incorrect.

If $n=1$ we have a simple $j$-pole leading to constant total cross section and vanishing elastic cross section. However such a behaviour of the cross sections is not supported by experimental data.

If $n=2$ we have the model of dipole pomeron ($\sigma_{t}(s)\propto
ln(s)$) and would like to emphasize that double $j$-pole is the maximal singularity of partial amplitude settled by unitarity bound
(\ref{eq:unitbound}) if its trajectory is linear at $t\approx
0$.

Thus, constructing the model leading to cross section which
increases faster than $ln(s)$, we need to consider a more
complicated case:
\begin{equation}\label{eq:j-gen**2}
\varphi (j,t)=\eta(j)\frac{\beta(j,t)}{\left
[j-1+k(-t)^{1/\mu}\right]^{n}}\approx \frac{i\beta(1,t)}{\left
[j-1+k(-t)^{1/\mu}\right]^{n}}.
\end{equation}
Making use of the same arguments as above, we obtain
\begin{equation}\label{eq:nmbound}
\sigma_{t}(s)\propto ln^{n-1}s,
\end{equation}
\begin{equation}
\sigma_{el}(s)\propto
ln^{2n-2-\mu}s \qquad \mbox{and} \qquad \mu\geq n-1. \nonumber
\end{equation}
However in this case amplitude $a(s,t)$ has a branch point at $t=0$ which is forbidden  by analyticity.

A proper form of amplitude leading to $t_{eff}$ \footnote{$t_{eff}$ can be defined by behaviour of elastic scattering amplitude at $s\to \infty $. If  $a(s,t)\approx sf(s)F(t/t_{eff}(s))$ then $\sigma_{el}(s)\propto|f(s)|^{2}\int_{-\infty}^{0}dt|F(t/t_{eff})|^{2}=
t_{eff}|f(s)F(1)|^{2}$.} decreasing faster than $1/lns$ (it is necessary for $\sigma_{t}$ rising faster than $lns$) is the following
\begin{equation}\label{eq:j-gencorrect}
\varphi (j,t)=\eta(j)\frac{\beta(j,t)}{\left
[(j-1)^{m}-kt\right]^{n}}.
\end{equation}
Now we have $m$ branch points colliding at $t=0$ in $j$-plane and creating the pole of order $mn$ at $j=1$ (but there is no branch point in $t$ at $t=0$). At the same time $t_{eff}\propto 1/ln^{m}s$ and from the $\sigma_{el}\propto ln^{2mn-2-m}s\leq \sigma_{t}\propto ln^{mn-1}s\leq
 ln^{2}s$ one can obtain
\begin{equation}\label{eq:mnbound}
\left \{
\begin{array}{ll}
mn &\leq m+1, \\
mn &\leq 3.
\end{array}
\right .
\end{equation}
If $\sigma_{el}\propto \sigma_{t}$ then $n=1+\frac{1}{m}$. Furthermore, if $\sigma_{t}\propto lns$ then $m=1$ and $n=2$ what corresponds just to the dipole pomeron model. In the tripole pomeron model $m=2$ and $n=3/2$ what means $\sigma_{t}\propto ln^{2}s$.

\section{Dipole parametrizations}\label{sec:Dpar}
The dominating term at high energy in this model is double pole
\begin{equation}\label{eq:dipole_d}
\varphi_{d}(j,t)\propto \frac{1}{(j-1-\alpha_{d}'t)^{2}}.
\end{equation}
Apparently in accordance with the inequalities (\ref{eq:mnbound}) the double pole obeys the unitarity limit for linear pomeron trajectory ($m=1$).
Adding to the partial amplitude less singular term (simple pole with a trajectory having intercept $\alpha(0)=1$ and a different slope $\alpha'$) we obtain dipole pomeron model in the form
\begin{equation}\label{eq:dipole_c}
\varphi(j,t)=\eta(j)\frac{\beta_{d}(t)}
{(j-1-\alpha_{d}'t)^{2}}+\eta(j)\frac{\beta_{s}(t)}
{j-1-\alpha_{s}'t}.
\end{equation}
It can be rewritten in $(s,t)$ representation as\begin{eqnarray}\label{eq:dipole_st}
a(s,t)=&&g_{d}ln(-iz)(-iz)^{1+\alpha_{d}'t}\exp(b_{d}t)\nonumber\\&&+
g_{s}(-iz)^{1+\alpha_{s}'t}\exp(b_{s}t),
\end{eqnarray}
where variable $z$ is proportional to cosine of scattering angle in
$t$-channel
\begin{equation}\label{eq:zvar}
 z=(t+2(s-2m_{p}^{2}))/z_{0}, \qquad z_{0}=1 {\rm GeV}^{2}.
\end{equation}
Generally, the form factors (or residues) $\beta(t)$ may be chosen in various forms (e.g. exponential, factorized powers {\it etc.}). However we consider the simplest exponential ones.

Let us consider two effective reggeons: crossing-even, $R_{+}(s,t)$, and crossing-odd, $R_{-}(s,t)$) instead of four contributions - $f, \omega$ and $\rho, a_{2}$ (the latter two reggeons are of less importance at high energy). We take into account their contribution in the standard form. However, we insert additional factor $Z_{R}(t)$ that changes a sign at some $t$ \footnote {For crossing-odd term of amplitude such a factor is well known and describes crossover effect, i.e. intersection of the $ab$ and $\bar ab$ differential cross sections at $t\approx -0.15$ GeV$^{2}$. Our analysis \cite{CLM} has shown that similar factor is visible in crossing-even reggeon term.}.
\begin{equation}\label{eq:reggeons}
R(s,t)=\eta_{R}g_{R}Z_{R}(t)(-iz)^{\alpha_{R}(t)}\exp(b_{R}t),
\end{equation}
where $\eta_{R}=-1/\sin(0.5\pi \alpha_{+}(0))$ for $R_{+}$-reggeon and $\eta_{R}=i/\cos(0.5\pi \alpha_{-}(0))$ for $R_{-}$-reggeon. Obviously these terms are very close to $f$- and $\omega$-reggeons, respectively. There are some arguments \cite{CLM} to use the factors $Z_{R}(t)$ in the form:
\begin{equation}\label{eq:Rzero}
Z_{R}(t)=\frac{\tanh(1+t/t_{R})}{\tanh(1)}.
\end{equation}
Going to extend wide regions of $s$ ($\sqrt{s}\geq 5$ GeV)
and $t$ ($0.1\leq |t|\leq 6$ GeV$^{2}$) \footnote{A more sophisticated
form for residues should be considered for larger
$|t|$.} we certainly need a few extra terms in amplitude to reach a good fit to the data. First of all it concerns the odderon contribution. The existing data on total cross section and parameters $\rho=\Re ea(s,0)/\Im ma(s,0)$, as  well known, do not show any visible odderon contribution. However, it appears definitely to provide the difference of $pp$ and $\bar pp$ differential cross sections at $\sqrt{s}=53$ GeV and $t$ around the dip. So, we add the odderon
contribution vanishing at $t=0$
\begin{eqnarray}\label{eq:odd_d}
{\cal O}(s,t)=t^{2}zZ_{R_{-}}(t)\biggl\{&& \!\!\!o_{1}ln^{2}(-iz)\exp(b_{o1}t)\nonumber\\
&&\!\!\!+o_{2}ln(-iz)\exp(b_{o2}t)\nonumber\\
&&\!\!\!+o_{3}\exp(b_{o3}t)\biggr\}(-iz)^{1+\alpha_{o}'t}
\end{eqnarray}
The term $\propto ln^{2}(s)$ in Eq. \ref{eq:odd_d} does not violate
unitarity restriction $\sigma_{el}(s)\leq \sigma_{t}(s)$ at very
large $s$ due to presence of factor $t^{2}$ (in the dipole model $t_{eff}\sim 1/lns $, therefore $\sigma_{el}\propto t^{3}_{eff}ln^{4}s\propto lns$).

At high energy and at $t=0$ two main rescattering terms of dipole pomeron (or cut terms) have the same form as the input amplitude - double pole plus simple pole. It means that comparing the model with experimental data we are not able to distinguish unambiguously input terms and cuts. Then as result, at $t=0$ one may use the input amplitude only. At $t\neq 0$ the situation occurs more complicated because the  slopes of trajectories in the cut terms are different from the input one. These terms are important at large $|t|$ but, in fact, they  are already taken into account at $t=0$.

Keeping in mind the above arguments and preserving a good description of the data at $t=0$ we take pomeron, pomeron-pomeron and pomeron-reggeons cuts vanishing at $t=0$. Certainly they are not "genuine" rescatterings but mimic them quite efficiently at $t\neq 0$. Thus we write down:

\noindent
the pomeron contribution
\begin{equation}\label{eq:pom}
P(s,t)=-g_{P}(-iz)^{1+\alpha_{P}'t}\left[\exp(b_{P1}t)-\exp(b_{P2}t)\right],
\end{equation}
the pomeron-pomeron cut
\begin{equation}\label{eq:Pcut}
C_{P}(s,t)=-\frac{t}{ln(-iz)}g_{PP}(-iz)^{1+\alpha_{P}'t/2}\exp(b_{PP}t),
\end{equation}
the pomeron-even reggeon cut
\begin{eqnarray}\label{eq:P+cut}
C_{R_{+}}(s,t)=&&-\frac{tZ_{R_{+}}(t)}{ln(-iz)}\eta_{R_{+}}
g_{P+}Z_{R_{+}}(t)\nonumber \\&&\times(-iz)^{\alpha_{+}(0)+\alpha'_{P+}t}\exp(b_{P+}t),
\end{eqnarray}
where
\begin{equation}\label{eq:slopeC+}
\alpha'_{P+}=\frac{\alpha_{P}'\alpha_{R_{+}}'}{\alpha_{P}'+\alpha_{R_{+}}'},
\end{equation}
and the pomeron-odd reggeon cut
\begin{eqnarray}\label{eq:P-cut}
C_{R_{-}}(s,t)=&&-i\frac{tZ_{R_{-}}(t)}{ln(-iz)}\eta_{R_{-}}
g_{P-}Z_{R_{-}}(t)\nonumber \\&&\times(-iz)^{\alpha_{-}(0)+\alpha'_{P-}t}\exp(b_{P-}t),
\end{eqnarray}
\begin{equation}\label{eq:slopeC-}
\alpha'_{P-}=\frac{\alpha_{P}'\alpha_{R_{-}}'}{\alpha_{P}'+\alpha_{R_{-}}'}.
\end{equation}

\section{Tripole pomeron model}\label{sec:Tpar}
As it follows from Eq.(\ref{eq:mnbound}) for the dominating
contribution in a tripole pomeron model with $\sigma_{t}(s)\propto
ln^{2}(s)$, i.e. $n=2$, $m=3/2$, we should take
\begin{equation}\label{eq:tripole1}
\varphi_{1}(j,t)=\eta(j)\frac{\beta_{1}(j,t)}{\left
[(j-1)^{2}-kt\right]^{3/2}}.
\end{equation}
It seems to be natural to write the subleading terms as the
following
\begin{equation}\label{eq:tripole2}
\varphi_{2}(j,t)=\eta(j)\frac{\beta_{2}(j,t)}{\left
[(j-1)^{2}-kt\right]},
\end{equation}
\begin{equation}\label{eq:tripole3}
\varphi_{3}(j,t)=\eta(j)\frac{\beta_{3}(j,t)}{\left
[(j-1)^{2}-kt\right]^{1/2}}.
\end{equation}
Then the amplitude has a form
\begin{equation}\label{eq:j-tripole}
\varphi
(j,t)=\varphi_{1}(j,t)+\varphi_{2}(j,t)+\varphi_{3}(j,t)+R(j,t),
\end{equation}
where $R(j,t)$ means the contribution of other reggeons and possible cuts
(which are important at low energies).

Taking into account that
\begin{equation}\label{eq:besselgen}
\int \limits_{0}^{\infty} dx x^{\alpha-1}e^{-\omega
x}\textit{J}_{\nu}(\omega_{0})=I_{\nu}^{\alpha}
\end{equation}
where
$$
\begin{array}{ll}
{I_{\nu}^{\nu+1}=\frac{(2\omega_{0})^{\nu}}{\sqrt{\pi}}
\frac{\Gamma(\nu+1/2)}
{(\omega^{2}+\omega_{0}^{2})^{\nu+1/2}}},\\ \\
{I_{\nu}^{\nu+2}=2\omega\frac{(2\omega_{0})^{\nu}}{\sqrt{\pi}}
\frac{\Gamma(\nu+3/2)}
{(\omega^{2}+\omega_{0}^{2})^{\nu+3/2}}},
\end{array}
$$
one can find
\begin{equation}\label{eq:phi1}
\frac{1}{(\omega^{2}+\omega_{0}^{2})^{3/2}}=\frac{1}{2\omega_{0}}
\int\limits_{0}^{\infty}
dx xe^{-x\omega}J_{1}(\omega_{0}x),
\end{equation}
\begin{equation}\label{eq:phi2}
\frac{1}{\omega^{2}+\omega_{0}^{2}}=\frac{1}{\omega_{0}}\int\limits_{0}^{\infty}
dx e^{-x\omega}\sin(x\omega_{0})
\end{equation}
and
\begin{equation}\label{eq:phi3}
\frac{1}{(\omega^{2}+\omega_{0}^{2})^{1/2}}=\int\limits_{0}^{\infty}
dx e^{-x\omega}J_{0}(\omega_{0}x).
\end{equation}
Thus tripole amplitude with the subleading terms can be presented as
\begin{eqnarray}\label{eq:trpole-st}
a_{tr}(s,t)=iz\biggl\{
&&g_{+1}\exp(b_{+1}t)ln(-iz)\frac{2J_{1}(\xi_{+}\tau_{+})}{\tau_{+}}\nonumber \\
&&+g_{+2}\frac{\sin(\xi_{+}\tau_{+})}{\tau_{+}}\exp(b_{+2}t)\nonumber\\
&&+g_{+3}J_{0}(\xi_{+}\tau_{+})\exp(b_{+3}t)\biggr \}
\end{eqnarray}
where $\xi_{+}=ln(-iz)+\lambda_{+}$, $z$ is defined by
Exp.(\ref{eq:zvar}), and $\tau_{+}=r_{+}\sqrt{-t/t_{0}}$, $t_{0}=1$
GeV$^{2}$, $r_{+}$ is a constant.

Similar expression for odderon contribution (but
introducing the factors $t$ and $Z_{R_{-}}(t)$) is given by
\begin{eqnarray}\label{eq:odd_tr}
{\cal O}(s,t)=ztZ_{R_{-}}(t)\biggl \{
&&g_{-1}ln(-iz)
\frac{2J_{1}(\xi_{-}\tau_{-})}{\tau_{-}}\exp(b_{-1}t)
\nonumber\\&&+g_{-2}
\frac{\sin(\xi_{-}\tau_{-})}{\tau_{-}}\exp(b_{-2}t)\nonumber\\&&+
g_{-3}J_{0}(\xi_{-}\tau_{-})\exp(b_{-3}t)\biggr \}
\end{eqnarray}
where $\xi_{-}=ln(-iz)+\lambda_{-}$ and
$\tau_{-}=r_{-}\sqrt{-t/t_{0}}$.

Again, similarly to the dipole model  we add the ``soft''
pomeron
\begin{equation}\label{eq:softpom}
P(s,t)=-g_{P}(-iz)^{1+\alpha_{P}'t}\exp(b_{P}t),
\end{equation}
the reggeon and cut contributions which are of the same form as
in dipole pomeron model
Eqs.(\ref{eq:Pcut},\ref{eq:P+cut},\ref{eq:P-cut}).

{\bf AGLN-model.} Let us give a few comments about another version of
tripole pomeron model presented in the papers \cite{GLN,AGN}.

1. If $\xi=ln(-is/s_{0}), s_{0}=1 {\rm GeV}^{2}$, then the first
pomeron term in \cite{GLN,AGN} is identical to the term in Eq. (\ref{eq:trpole-st})
while for the second and third terms authors use
$$
g_{2}(t)\xi J_{0}(\xi\tau)
$$
and
$$
g_{3}(t)[J_{0}(\xi\tau_{+})-\xi_{0}\tau_{+} J_{1}(\xi\tau_{+})]
$$
which originated from the partial amplitudes
$$
\varphi(j,t)=\eta(j)g_{2}(t)\frac{2(j-1)}{\left
[(j-1)^{2}-kt\right]^{3/2}},
$$
and
$$
\varphi(j,t)=\eta(j)g_{2}(t)\frac{(j-1)^{2}+kt}{\left
[(j-1)^{2}-kt\right]^{3/2}},
$$
respectively.

2. They used another form of odderon terms. The maximal odderon
contribution in the form
\begin{eqnarray}\label{eq:maxodd}
{\cal
O}_{m}(s,t)&&=g_{-1}ln^{2}(-iz)\frac{\sin(\xi\tau_{-})}{\tau_{-}}\exp(b_{-1}t)
\nonumber\\&&+
g_{-2}ln(-iz)\cos(\xi\tau_{-})\exp(b_{-2}t)\nonumber\\&&+g_{-3}\exp(b_{-3}t)
\end{eqnarray}
as well as a simple pole odderon and odderon-pomeron cut are also taken
into account.

3. Omitting the details we note that because of the chosen form of signature factors the AGLN amplitude has pole in physical region at $t=-1/\alpha'=-4$ GeV$^{2}$. This feature of the model restricts its applicability region. AGLN amplitude has similar poles even at lower values of $|t|$ in the reggeon terms. Thus the model requires a modification to describe wider region of $t$ than was considered in \cite{AGN}, namely $|t|\leq 2.6$ GeV$^{2}$.

4. Clearly this model leads to the unreasonable intercept value for the
crossing-odd reggeon, $\alpha_{-}(0)=0.34$. It is in strong contradiction with the values known from meson resonance spectroscopy data. One could expect it close to the intercept of $\omega$-trajectory, $\alpha_{\omega}(0)\approx 0.43 - 0.46$ \cite{DGMP2}.

Nevertheless, in the Section \ref{sec:expdata} we demonstrate the curves for differential cross sections obtained in AGLN model in comparison with the results of our dipole and tripole models at energies available and future LHC.

\section{Comparison with experimental data}\label{sec:expdata}
\subsection{Total cross sections}
Analyzing the $pp$ and $\bar pp$ data we keep in mind a further extension of the models to elastic $\pi^{\pm} p$ and $K\pm p$ scattering, which are quite precisely measured. One important point should be underlined from the beginning. Fitting the $pp$ and $\bar pp$  data  on $\sigma_{t}$ and $\rho$ gives the set of parameters which is essentially different from those derived from the fit of all ($p, \bar p, \pi$- and $K$-meson) the data.

Following this procedure at the first stage we determine all parameters which control the amplitudes at $t=0$. We use the standard data set for the $\pi^{\pm} p$ and $K\pm p$ total cross sections and the ratios $\rho$ (at 5 GeV$\leq\sqrt{s}<$2000 GeV) \cite{PDG} to find intercepts of $C_{\pm}$-reggeons and couplings of the reggeon and pomeron exchanges. There are 542 experimental points in the  region under consideration (see Table \ref{tab:t0-1}).

An extension of the $pp\rightarrow pp$ and $\bar pp\rightarrow \bar pp$ dipole and tripole amplitudes to $\pi^{\pm} p$ and $K^{\pm} p$ elastic scattering is quite straight forward. All the couplings are various in these amplitudes at $t=0$ but the odderon does not contribute to the $\pi p$ and $Kp$ amplitudes. In the simplest unitarization schemes (eikonal, $U$-matrix) all total cross
sections at asymptotically high energies have the universal behaviour, $\sigma_{t}(s)\to \sigma_{0}\log^{2}(s_{ap}/s_{0})$, where $\sigma_{0}$ is independent of the initial particles. Today the data available support this conclusion and advocates putting the same couplings $g_{+1}^{p}=g_{+1}^{\pi}=g_{+1}^{K}$ for the leading pomeron terms in all amplitudes. Besides, in order to avoid uncertainty  at $t=0$ (constant contributions to total cross sections come additively from the third term of Eq.(\ref{eq:trpole-st}) and from ``soft'' pomeron, Eq.(\ref{eq:softpom})) we substitute couplings $g_{+3}$ for $g_{+3}-g_{P}$ in Eq.(\ref{eq:trpole-st}). As a result we have energy independent contribution to the total cross from $g_{+3}$ only.

The following normalization of $ab\rightarrow ab$ amplitude is used
\begin{equation}\label{eq:norm}
\sigma_t=\frac{1}{s_{ab}}\Im mA(s,0), \qquad
\frac{d\sigma}{dt}=\frac{1}{16\pi s_{ab}^{2}}|A(s,t)|^{2}
\end{equation}
where
$$
s_{ab}=\sqrt{(s-m_a^2-m_b^2)^2-4m_a^2m_b^2}=2p_a^{lab}\sqrt{s}
$$
and $p_a^{lab}$ is the momentum of hadron $a$ in laboratory system of $b$.

The details of the fit at $t=0$ are presented in the Tables \ref{tab:t0-1} and \ref{tab:t0-2}.

\begin{table}
  \caption{Quality of the fit to $\sigma_{t}$ and $\rho$}\label{tab:t0-1}
\begin{ruledtabular}
\begin{tabular}{cccc}
\multicolumn{2}{c}{} & \multicolumn{2}{c}{$\chi^{2}_{tot}/N_{p}$ }\\
\hline
 quantity & number of data & Dipole model & Tripole model \\
\hline
$\sigma_{t}^{pp}$ & 104        & 0.88260E+00 & 0.87055E+00 \\
 $\sigma_{t}^{\bar pp}$ &  59  & 0.95280E+00 & 0.96273E+00 \\
 $\sigma_{t}^{\pi^{+} p}$ & 50 & 0.66216E+00 & 0.66792E+00 \\
 $\sigma_{t}^{\pi^{-}p}$ &  95 & 0.10023E+01 & 0.99864E+00 \\
 $\sigma_{t}^{K^{+}p}$ &  40   & 0.72357E+00 & 0.72104E+00 \\
 $\sigma_{t}^{K^{-}p}$ &  63   & 0.61392E+00 & 0.60883E+00 \\
 $\rho^{pp}$ & 64              & 0.16612E+01 & 0.16965E+01 \\
 $\rho^{\bar pp}$ &  11        & 0.40392E+00 & 0.40675E+00 \\
 $\rho^{\pi^{+} p}$ &  8       & 0.15107E+01 & 0.15036E+01 \\
 $\rho^{\pi^{-}p}$ &  30       & 0.12560E+01 & 0.12122E+01 \\
 $\rho^{K^{+}p}$ &  10         & 0.10869E+01 & 0.10016E+01 \\
 $\rho^{K^{-}p}$ &  8          & 0.12185E+01 & 0.11611E+01 \\
\hline
 & & \multicolumn{2}{c}{$\chi^{2}_{tot}/dof$ }\\
Total  &  542         & 0.99450E+00 & 0.99345E+00 \\
\end{tabular}
\end{ruledtabular}
\end{table}

\begin{table*}
  \centering
  \caption{Intercepts and couplings (GeV$^{-2}$) in the Dipole and Tripole
  models from the fit to $\sigma_{t}$ and $\rho$}
  \label{tab:t0-2}
\begin{ruledtabular}
\begin{tabular}{lrrlrr}
\multicolumn{3}{c}{Dipole model} & \multicolumn{3}{c}{Tripole model}  \\
\hline
 {\small parameter} & value & error & {\small parameter} & value & error \\
\hline
 $\alpha_{R+}(0)$  &  0.80846E+00 & 0.36035E-02 &  $\alpha_{R+}(0)$  & 0.71947E+00 & 0.18496E-02\\
 $\alpha_{R-}(0)$  &  0.46505E+00 & 0.91416E-02 &  $\alpha_{R-}(0)$  & 0.46356E+00 & 0.73746E-02\\
 $g_{d}^{p}$       &  0.89435E+01 & 0.19499E+00 &  $g_{+1}^{p}$      & 0.15330E+02 & 0.31619E+00\\
 $g_{s}^{p}$       & -0.52159E+02 & 0.30078E+01 &  $g_{+2}^{p}$      & 0.19153E+01 & 0.38589E-01\\
 $g_{R+}^{p}$      &  0.15857E+03 & 0.30846E+01 &  $g_{+3}^{p}$      & 0.20672E+00 & 0.26747E-02\\
 $g_{R-}^{p}$      &  0.58961E+02 & 0.28775E+01 &  $g_{R+}^{p}$      & 0.96906E+02 & 0.84678E+00\\
 $g_{d}^{\pi}$     &  0.70477E+01 & 0.22719E+00 &  $g_{R-}^{p}$      & 0.59294E+02 & 0.23228E+01\\
 $g_{s}^{\pi}$     & -0.45720E+02 & 0.29022E+01 &  $g_{+1}^{\pi}$    & 0.69901E+01 & 0.18396E+00\\
 $g_{R+}^{\pi}$    &  0.10691E+03 & 0.33590E+01 &  $g_{+2}^{\pi}$    & 0.10106E+01 & 0.27123E-01\\
 $g_{R-}^{\pi}$    &  0.10710E+02 & 0.52507E+00 &  $g_{R+}^{\pi}$    & 0.56296E+02 & 0.45360E+00\\
 $g_{d}^{K}$       &  0.54351E+01 & 0.29205E+00 &  $g_{R-}^{\pi}$    & 0.10784E+02 & 0.43687E+00\\
 $g_{s}^{K}$       & -0.29703E+02 & 0.33234E+01 &  $g_{+1}^{K}$      & 0.12186E+02 & 0.22424E+00\\
 $g_{R+}^{K}$      &  0.70785E+02 & 0.43792E+01 &  $g_{+2}^{K}$      & 0.14683E+00 & 0.30439E-01\\
 $g_{R-}^{K}$      &  0.23674E+02 & 0.11159E+01 &  $g_{R+}^{K}$      & 0.28730E+02 & 0.51047E+00\\
                   &                &               &  $g_{R-}^{K}$  & 0.23831E+02 & 0.91697E+00\\
\end{tabular}
\end{ruledtabular}
\end{table*}

\subsection{Differential cross sections}  At the second stage of the
fitting procedure we fix all the intercept and coupling values obtained at
the first stage. The other parameters are determined by fitting the $d\sigma/dt$ data in the region
\begin{equation}\label{eq:t-region}
0.1 {\rm GeV}^{2}\leq |t|\leq 6 {\rm GeV}^{2}, \quad \sqrt{s}\geq 5
{\rm GeV}.
\end{equation}

The measurements of the differential elastic cross sections were very intensive for the last 40 years. Fortunately, most of them have been collected in the Durham Data Base \cite{DDB}. However, there are 80 papers, with different conventions, and various units. The complete list of the references is given by \cite{CLM}. We have uniformly formatted them, found and corrected some errors in the sets and gave a detailed description of a full set which contains about 10000 points. Analyzing each subset of these data we have payed \cite{CLM}  particular attention to the data at small $t$. Some of subsets which are in strong disagreement with the rest of the dataset were excluded from the fit.

A similar work has been done for the data at $|t|>0.7$ GeV$^{2}$. We have found out and corrected some mistakes in the data base. Furtermore, we excluded from the final dataset the subsets \cite{BRUNETON} at $\sqrt{s}=9.235$ GeV, \cite{CONETTI} at $\sqrt{s}=19.47$ and $27.43$ GeV from $pp$ data and \cite{BOGOLYUBSKY} at $\sqrt{s}=7.875$ GeV, \cite{AKERLOF} at $\sqrt{s}=9.778$ GeV from $\bar pp$ data because they strongly contradict the bulk of data. Thus, the considered models were fitted to 2532 points of $d\sigma/dt$ in the region Eq.(\ref{eq:t-region}). The results are given in the Table \ref{tab:chi2-t} for a quality of fitting and in the Table \ref{tab:par-t} for the fitting parameters.

\begin{figure}[h]
\includegraphics[scale=0.35]{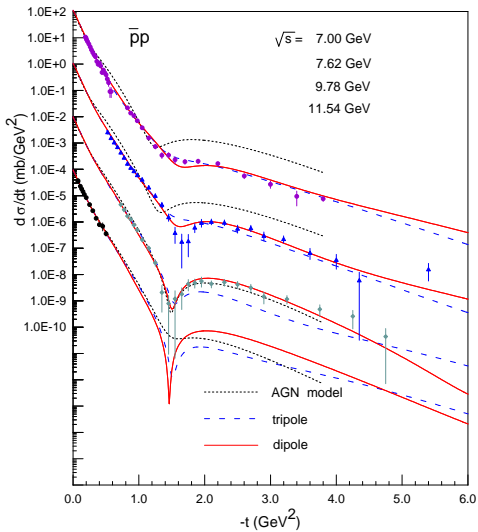}
\caption{$\bar pp$ at low energies}
\label{fig:paplow}
\end{figure}

\begin{figure}[b]
\includegraphics[scale=0.35]{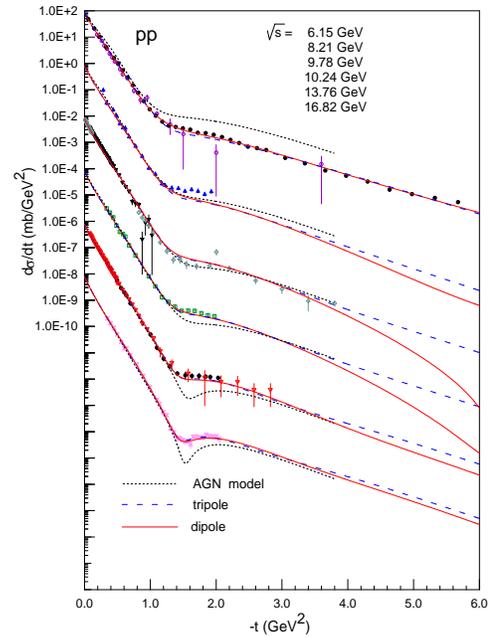}
\caption{$ pp$ at low energies}
\label{fig:pplow}
\end{figure}

\begin{figure}[h]
\includegraphics[scale=0.35]{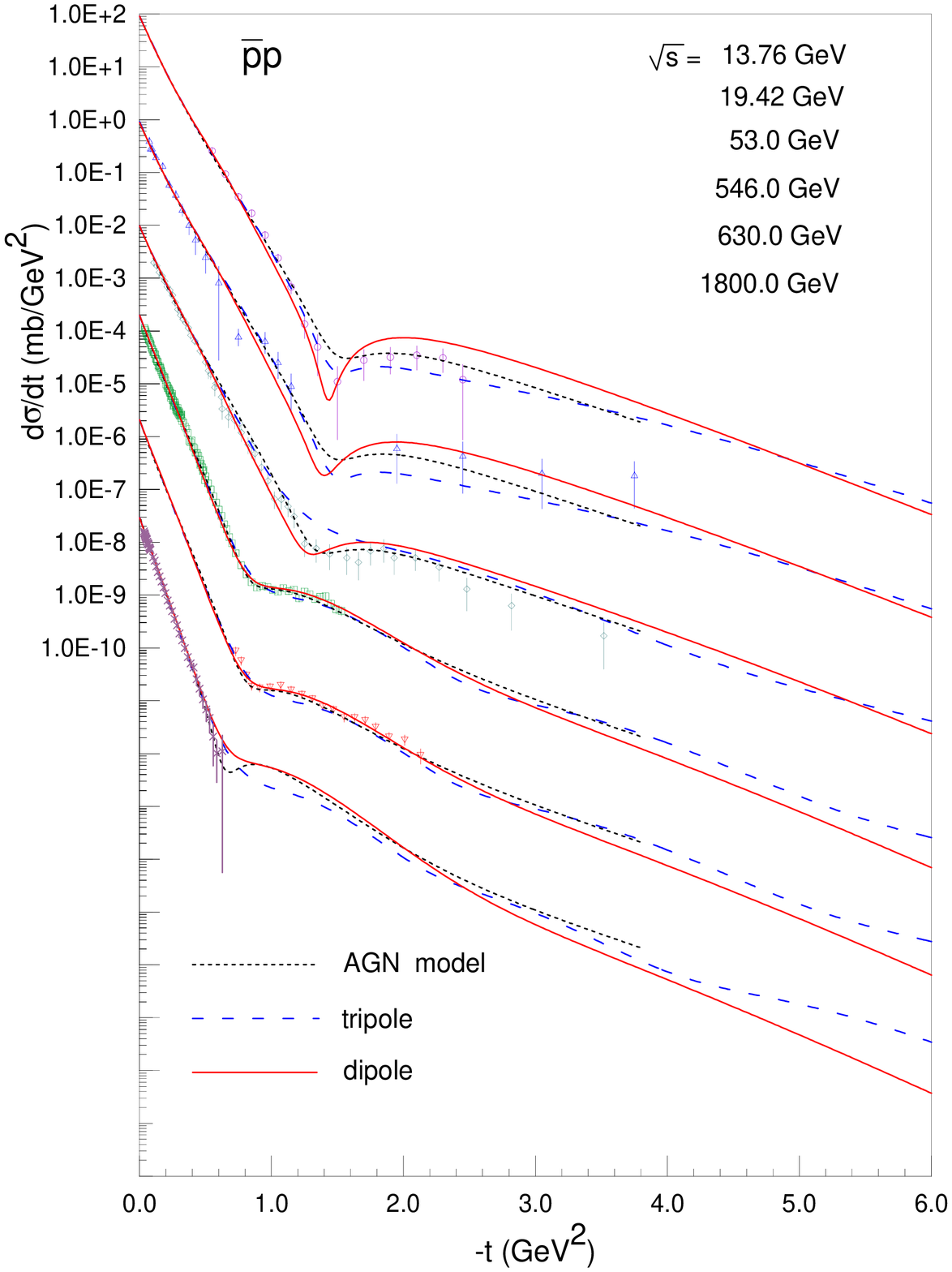}
\caption{$\bar pp$ at high energies}
\label{fig:paphigh}
\end{figure}

\begin{figure}[b]
\includegraphics[scale=0.35]{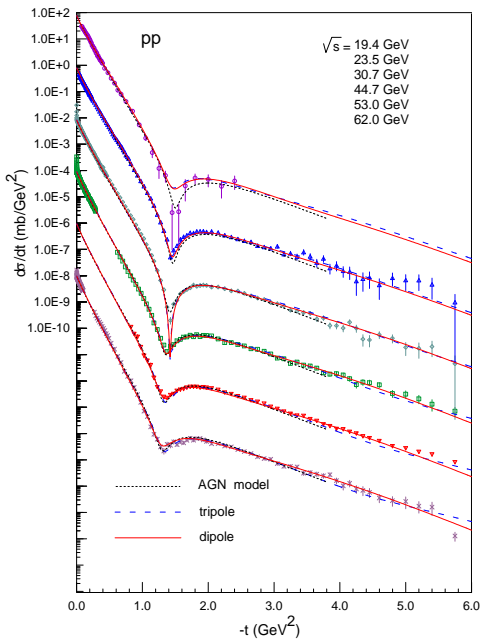}
\caption{$pp$ at high energies}
\label{fig:pphigh}
\end{figure}

\begin{figure}[b]
\includegraphics[scale=0.5]{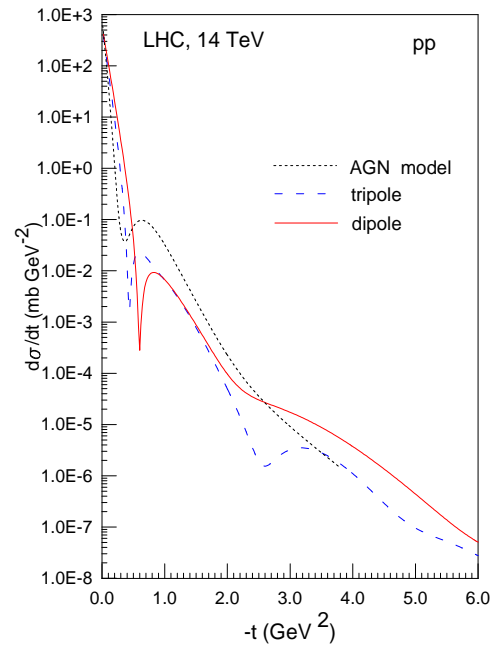}
\caption{Predictions for LHC energy}
\label{fig:lhc}
\end{figure}

In the Figs. \ref{fig:paplow} - \ref{fig:lhc} we show experimental data at some energies and theoretical curves obtained in three models: AGLN \cite{AGN}, Dipole and Tripole. As to the AGLN model, we would like to emphasize that the corresponding curves were calculated at the parameters given in \cite{AGN}. However, in contrast to Dipole and Tripole models AGLN model was fitted to differential cross sections at $\sqrt{s}>9.7$ GeV and $|t|<2.6$ GeV$^{2}$ but not with a complete dataset. Therefore a disagreement between curves and data behaviours at lowest energies is not surprising in the given model. The AGLN model works well at high energies.

\smallskip

~

\begin{table}
  \caption{Quality of the fit to $d\sigma/dt$}\label{tab:chi2-t}
\begin{ruledtabular}
\begin{tabular}{cccc}
  & Number of &
\multicolumn{2}{c}{$\chi^{2}_{tot}/N_{p}$}\\
\cline{3-4}
&points, $N_{p}$ & Dipole model &
Tripole model\\
\hline
 $d\sigma^{pp}/dt$      & 1857 & 0.15122E+01 & 0.18153E+01
\\
 $d\sigma^{\bar pp}/dt$ & 675  & 0.14183E+01 & 0.16697E+01\\
\end{tabular}
\end{ruledtabular}
\end{table}

\begin{table*}
  \caption{Parameters of the models, from the fit to $d\sigma/dt$
  (parameters $r_{\pm}, \lambda_{\pm}$ are dimensionless, $t_{R\pm}$
  are given in GeV$^{2}$, the rest parameters are given in GeV$^{-2}$).}
  \label{tab:par-t}
\begin{ruledtabular}
 \begin{tabular}{lrrlrr}
\multicolumn{3}{c}{Dipole model} &
\multicolumn{3}{c}{Tripole model}\\
 parameter & value & error & parameter & value & error\\
$\alpha_{d}'$  & 0.30631E+00 & 0.16923E-02 &$r_{+}$         & 0.25417E+00 & 0.33181E-02  \\
$\alpha_{s}'$  & 0.28069E+00 & 0.19026E-03 &$\lambda_{+}$   & 0.11575E+01 & 0.14824E+00  \\
$b_{d}$        & 0.38675E+01 & 0.22767E-01 &$b_{+1}$        & 0.34583E+01 & 0.37982E-01  \\
$b_{s}$        & 0.55679E+00 & 0.14694E-02 &$b_{+2}$        & 0.19091E+01 & 0.30598E-01  \\
$\alpha_{R+}'$ & 0.82000E+00 & fixed        &$b_{+3}$       & 0.45970E+00 & 0.29750E-02  \\
$b_{R+}$       & 0.29226E+01 & 0.30019E-01 &$\alpha_{R+}'$  & 0.82000E+00 & fixed       \\
$t_{R+}$       & 0.48852E+00 & 0.26683E-02 &$b_{R+}$        & 0.10668E+01 & 0.30350E-01  \\
$\alpha_{R-}'$ & 0.91000E+00 & fixed        &$t_{R+}$       & 0.54237E+00 & 0.13817E-01  \\
$b_{R-}$       & 0.15201E+01 & 0.68671E-01 &$\alpha_{R-}'$  & 0.91000E+00 & fixed       \\
$t_{R-}$       & 0.14497E+00 & 0.24811E-02 &$b_{R-}$        & 0.61435E-01 & 0.20044E-01  \\
$o_{1}$        & 0.30738E+00 & 0.31368E-02 &$t_{R-}$        & 0.15755E+00 & 0.26068E-02  \\
$o_{2}$        &-0.63119E+01 & 0.55282E-01 &$r_{-}$         & 0.78807E-01 & 0.60563E-02  \\
$o_{3}$        & 0.13456E+00 & 0.21551E-02 &$\lambda_{-}$   & 0.16281E+02 & 0.19020E+01  \\
$\alpha_{o}'$  & 0.21810E-01 & 0.70324E-03 &$o_{1}$         &-0.56075E-01 & 0.51702E-02  \\
$b_{o1}$       & 0.39317E+01 & 0.81697E-02 &$o_{2}$         & 0.17372E+01 & 0.17893E+00  \\
$b_{o2}$       & 0.45007E+01 & 0.76861E-02 &$o_{3}$         &-0.61193E+02 & 0.36330E+01  \\
$b_{o3}$       & 0.12947E+01 & 0.76773E-02 &$b_{o1}$        & 0.12038E+01 & 0.26425E-01  \\
$g_{P}$        & 0.58961E+02 & 0.98576E-01 &$b_{o2}$        & 0.15152E+01 & 0.31722E-01  \\
$\alpha_{P}'$  & 0.30696E+00 & 0.20475E-03 &$b_{o3}$        & 0.26331E+01 & 0.78349E-01  \\
$b_{P1}$       & 0.54894E+00 & 0.15036E-02 &$g_{P}$         & 0.16042E+02 & 13856E-01  \\
$b_{P2}$       & 0.59365E+01 & 0.34863E-01 &$\alpha_{P}'$   & 0.36060E+00 & 0.76335E-02  \\
$g_{PP}$       &-0.39324E+02 & 0.36883E+00 &$b_{P}$         & 0.14662E+01 & 0.29152E-01  \\
$b_{PP}$       & 0.11828E+01 & 0.44025E-02 &$g_{PP}$        & 0.91195E+01 & 0.84044E+00  \\
$g_{P+}$       &-0.22656E+03 & 0.27529E+01 &$b_{PP}$        & 0.44977E+00 & 0.46404E-01  \\
$b_{P+}$       & 0.17522E+01 & 0.98459E-02 &$g_{P+}$        & 0.11772E+02 & 0.57753E+00  \\
$g_{P-}$       &-0.15255E+02 & 0.28213E+00 &$b_{P+}$        & 0.81585E-01 & 0.28185E-01  \\
$b_{P-}$       & 0.24068E-01 & 0.61336E-02 &$g_{P-}$        & 0.81908E+01 & 0.88408E+00  \\
               &             &             &$b_{P-}$        &-0.79115E-01 & 0.45833E-01  \\
\end{tabular}
\end{ruledtabular}
\end{table*}

\section{Conclusion}
In this paper we compare three unitarized models of elastic scattering amplitude fitting the Dipole and Tripole models to all existing data. We emphasize that the amplitude leading to the behaviour of $\sigma_{t}\propto \ln^{2}s$ should be parameterized with a special care of the unitarity and analyticity restrictions on properties of the leading partial wave singularity.

The Figures and Tables demonstrate good description of the data within the considered models. However the obtained $\chi^{2}$ (Table \ref{tab:chi2-t}) hints that the Dipole pomeron model looks more preferable.

We believe the most interesting and instructive result for further search of more realistic model is shown in Fig.~\ref{fig:lhc}. Our predictions of the compared models (together with AGLN model) for $pp$ cross section at LHC energy are crucially different at $|t|$ around 0.3 - 0.5 GeV$^{2}$. Certainly the future TOTEM measurement will allow to distinguish between three considered models.

I would like to thank Prof.~B.~Nicolescu and Dr.~J.R.~Cudell for many useful discussions.

The work was supported partially by the Ukrainian Fund of Fundamental Researches.

\end{document}